\documentclass[aps,prd,twocolumn,floatfix,nofootinbib]{revtex4-2}
\usepackage{graphicx}
\usepackage{amsmath,amssymb,mathtools,siunitx,physics,hyperref}
\usepackage{color}

\begin{document}
\title{Finite signal-to-noise ratio bias in parameter estimation for damped oscillations: cautionary remark about catalog-level black-hole spectroscopy}
\author{Koutarou Kyutoku$^{1,2}$}
\affiliation{$^1$Department of Physics, Graduate School of Science, Chiba University, Chiba 263-8522, Japan}
\affiliation{$^2$Interdisciplinary Theoretical and Mathematical Sciences Program (iTHEMS), RIKEN, Wako, Saitama 351-0198, Japan}

\date{\today}

\begin{abstract}
 We investigate biases in parameter estimation for damped oscillations motivated by applications to black-hole spectroscopy in gravitational-wave physics. Focusing on the simplest model of a single-mode damped sinusoid with a fixed start time in white noise, we show that, at finite signal-to-noise ratio $\rho$, the damping time is biased toward larger values without being suppressed by the quality factor for two reasons. One is the gradient of the prior, through which the damping time is affected by the typically decreasing prior on the amplitude. The other is a higher-order finite-$\rho$ correction to the likelihood geometry. These biases arise even if the model and analysis are appropriate. Moreover, they could be exaggerated in naive joint inferences from catalog events. Quantitatively, if estimates from multiple events with $\rho = 10$ are combined without due care, the catalog-level black-hole spectroscopy could report false violation of the Kerr hypothesis with $\gtrsim 100$ events. We also propose simple strategies to mitigate these biases at the level of individual events.
\end{abstract}

\maketitle

\section{Introduction} \label{sec:intro}

Black-hole spectroscopy is one of the most active research areas in current gravitational-wave physics \cite{2004CQGra..21..787D} (see Ref.~\cite{2025arXiv250523895B} for reviews). Because general relativity states that an astrophysical black hole in a vacuum must be a Kerr black hole characterized only by the mass and spin, its perturbative response is also determined by these two parameters. Thus, whether a putative black hole is consistent with the Kerr hypothesis can be tested by measuring the spectrum of quasinormal modes specified by the frequency $f$ and the damping time $\tau$ (see Ref.~\cite{2009CQGra..26p3001B} for reviews). As the number of detected binary black holes increases rapidly, the black-hole spectroscopy has actively been pursued in this decade \cite{2016PhRvL.116v1101A,2021PhRvD.103l2002A,2025PhRvD.112h4080A,2026arXiv260319021T}. While various implementations of this program have been developed, one powerful method is to compare $f$ and $\tau$ of the dominant, $l=m=2$ fundamental mode estimated from the ringdown phase with their values predicted from the inspiral phase assuming general relativity. For example, pSEOBNR analysis performed by LIGO-Virgo-KAGRA collaborations belongs to this category \cite{2021PhRvD.103l2002A,2025PhRvD.112h4080A,2026arXiv260319021T}.

Recent analysis begins to suggest nonnegligible deviations of the estimated posteriors from the prediction for Kerr black holes \cite{2026arXiv260319021T}. In particular, the estimated damping time is suggested to be longer than the prediction. Table~3 of Ref.~\cite{2026arXiv260319021T} reports that the posterior median shows positive deviations for 15 out of 19 signals in pSEOBNR analysis, and this occurs only $<1\%$ if the signature is random. Accordingly, when the posteriors are combined across the catalog, the Kerr hypothesis tends to be disfavored (see Fig.~4 of Ref.~\cite{2026arXiv260319021T}), although this may not be as problematic as it appears \cite{2024PhRvD.109h1302P}. If these deviations are a genuine feature of astrophysical signals, it indicates the breakdown of the hypothesis that the merger remnants are vacuum black holes in general relativity. If this is true, the possible resolution may include that the remnant black hole is surrounded by known or unknown matter, the remnant is not a black hole, or general relativity is violated.

Before declaring that the merger remnant is not a Kerr black hole, any possible biases inherent in the black-hole spectroscopy need to be examined with the utmost care. While previous studies have investigated biases caused by imperfect models and/or improper analysis (see, e.g., Refs.~\cite{2017PhRvD..96j2004T,2018PhRvD..97d4048B,2021arXiv210705609I,2025PhRvD.112h4076V}), biases are also caused by the finite signal-to-noise ratio $\rho$ even if both the model and analysis are appropriate. Specifically, as we demonstrate below, each parameter suffers an $\order{\rho^{-2}}$ bias with basically the same signature \cite{2008PhRvD..77d2001V}. Once $N$ events in a catalog are combined without due care, the systematic bias surpasses the statistical error suppressed by $\sim N^{-1/2}$. That is, catalog-level black-hole spectroscopy could erroneously reject the Kerr hypothesis even with the perfect model and proper analysis for individual events, unless these biases are addressed appropriately. We recall that the ringdown signals in individual events have not been very loud so far. For example, pSEOBNR analysis requires the ringdown to have only $\rho \ge 8$ \cite{2026arXiv260319021T}.

In this study, we investigate the biases caused by the finite signal-to-noise ratio \cite{1994PhRvD..49.2658C,2008PhRvD..77d2001V}. To avoid being too specific to a particular pipeline, we consider a general problem of parameter estimation for damped oscillations by matched filtering (see, e.g., Refs.~\cite{2004eusi.conf...73D,2007PhRvD..75l4017B} for studies in the context of linear signal processing). Accordingly, while we mainly focus on the black-hole spectroscopy, our results will be applicable in a broad context. We show that the damping time is most susceptible to biases in the sense that its bias is not suppressed by the quality factor $Q \coloneqq \pi f \tau$ of damped oscillations, here the quasinormal modes of black holes \cite{2006PhRvD..73f4030B}. We also propose simple strategies to mitigate these biases for individual events.

This paper is organized as follows. We summarize the waveform model and assumptions, analysis methods, and statistical errors in Sec.~\ref{sec:setup}. The biases are discussed in Sec.~\ref{sec:pg} regarding the gradient of the prior and in Sec.~\ref{sec:lg} regarding the geometry of likelihood surface. Section \ref{sec:summary} is devoted to a summary and discussion.

\section{Setup and statistical error} \label{sec:setup}

We adopt a damped-sinusoid waveform model with four parameters $\{A , \phi , f , \tau\}$,
\begin{equation}
 h(t) = A e^{-t/\tau} \cos (2\pi f t + \phi) \Theta (t) ,
\end{equation}
where $\Theta (t)$ is Heaviside's step function. We always fix the start time of data analysis, here $t=0$, to focus on the biases that emerge even in the absence of mismodeling. In reality, an appropriate choice of the start time is a subtle issue in black-hole spectroscopy (see, e.g., Refs.~\cite{2020PhRvD.101d4033B,2023PhRvD.108j4020B,2024CQGra..41s5023T,2024PhRvD.109l4030C}) due to complications associated with overtones, higher harmonics, and nonlinear modes as well as the prompt and tail contributions \cite{1986PhRvD..34..384L}. Future extensions to including these additional components, particularly multiple modes, will be beneficial. We leave this task for future study, expecting that these extensions will not conspire to reduce the biases discussed below.

\begin{figure}[tbp]
 \includegraphics[width=.95\linewidth]{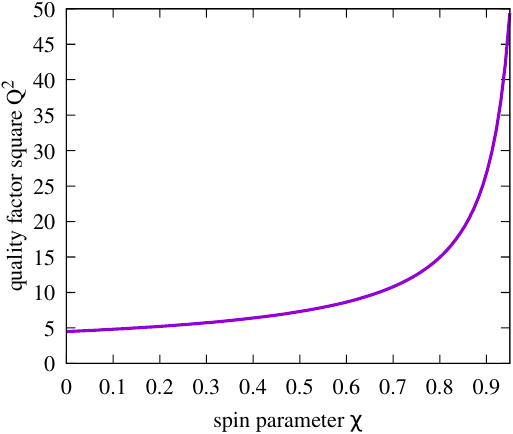}
 \caption{Square of the quality factor $Q = \pi f \tau$ for the $l=m=2$ fundamental quasinormal mode of Kerr black holes as a function of the spin parameter in the range $0 \le \chi \le 0.95$ calculated using the fitting formula of Ref.~\cite{2006PhRvD..73f4030B}. For $\chi \approx 0.686$ resulting from mergers of equal-mass, nonspinning black holes \cite{2009PhRvD..79b4003S}, $Q^2 \approx 10$.} \label{fig:qfactor}
\end{figure}

We evaluate the noise-weighted inner product of two waveforms in the time domain assuming that the power spectral density of the detector noise is approximately constant in the frequency range relevant to the damped sinusoid. That is, the inner product between two waveforms $a(t)$ and $b(t)$, both starting at $t=0$, is defined by
\begin{equation}
 (a|b) \coloneqq \frac{2}{S_n} \int_0^\infty a(t) b(t) \dd{t} ,
\end{equation}
where $S_n$ is the one-sided power spectral density around the frequency $f$ of the sinusoid. We also assume that the detector noise is stationary and Gaussian for simplicity, as commonly done in the literature. Realistic treatment is unlikely to mitigate the biases. The norm of $h(t)$ gives the optimal signal-to-noise ratio as
\begin{equation}
 \rho^2 \coloneqq (h|h) = \frac{A^2 \tau}{2S_n} \pqty{1 + \frac{\cos 2\phi - 2Q\sin 2\phi}{4Q^2+1}} ,
\end{equation}
where $Q = \pi f \tau$ is the quality factor \cite{1989PhRvD..40.3194E,1992PhRvD..46.5236F}. When quasinormal modes with complex frequency $\omega = \omega_\mathrm{R} + \mathrm{i} \omega_\mathrm{I}$ are considered , $Q = \omega_\mathrm{R}/(2\omega_\mathrm{I})$. Figure \ref{fig:qfactor} shows $Q^2$ for the $l=m=2$ fundamental mode of Kerr black holes adopting the fitting formula of Ref.~\cite{2006PhRvD..73f4030B}.

Throughout this study, we employ the (beyond-)Fisher approximation, in which the waveform is expanded in terms of $\rho^{-1}$ around the true value \cite{1992PhRvD..46.5236F,1994PhRvD..49.2658C,2008PhRvD..77d2001V}. Statistical errors (in an appropriate meaning for specific analysis) are evaluated via the Fisher matrix $\Gamma_{ab} \coloneqq (h_a|h_b)$, where $h_a \coloneqq \partial_a h$ denotes the derivative of $h(t)$ with respect to the waveform parameters $\theta^a = \{A , \phi , f , \tau\}$. In this study, lowercase Latin letters $a$--$d$ indicate these parameters, and repeated up-down indices are summed over. We avoid choosing any variable in logarithms to simplify the computation of biases. Because the initial phase $\phi$ is not of interest in typical black-hole spectroscopy, we mainly discuss quantities averaged (or marginalized) over $\phi$, denoted by brackets as $\ev{X} \coloneqq \int_0^{2\pi} X(\phi) \dd{\phi} / (2\pi)$. The amplitude $A$ can be eliminated by expressing quantities in terms of $\ev{\rho^2} = A^2 \tau /(2S_n)$.

The Fisher matrix can be computed and inverted to derive the covariance matrix $C^{ab}=(\Gamma^{-1})^{ab}$ in a straightforward manner \cite{2006PhRvD..73f4030B}. As perhaps physically expected, the determinant of the Fisher matrix,
\begin{equation}
 \det\,(\Gamma_{ab}) = \ev{\rho^2}^4 \frac{64\pi^2 Q^8}{A^2 (4Q^2 + 1)^4} ,
\end{equation}
does not depend on $\phi$. Thus, $C^{ab}$ depends on $\phi$ only via the numerators, and we may readily average (or marginalize) over $\phi$ to obtain the averaged covariance matrix $\ev{C^{ab}}$.

The variance for each parameter is given by the diagonal elements as $\sigma_a^2 = C^{aa}$. We choose to average these quantities after normalizing by $\rho^2$, and this derives
\begin{align}
 \frac{\ev{\rho^2 \sigma_A^2}}{A^2} & = 2 + \frac{3}{2Q^2} + \frac{7}{16Q^4} + \frac{1}{32Q^6} , \\
 \ev{\rho^2 \sigma_\phi^2} & = 2 + \frac{5}{2Q^2} + \frac{13}{16Q^4} + \frac{3}{32Q^6} , \\
 \frac{\ev{\rho^2 \sigma_f^2}}{f^2} & = \frac{1}{Q^2} + \frac{5}{8Q^4} + \frac{1}{32Q^6} , \\
 \frac{\ev{\rho^2 \sigma_\tau^2}}{\tau^2} & = 4 + \frac{3}{2Q^2} + \frac{3}{8Q^4} .
\end{align}
This averaging procedure is not unique, and we may equally average $\sigma_a^2$ themselves fixing waveform parameters. Fortunately, the leading-in-$Q$ dependence is unaffected as shown in App.~\ref{app:directave} (see also below). Considering that $A$ and $\tau$ should depend on various extrinsic and intrinsic parameters in reality, we expect that our average that respects directly observable values $\rho$ is reasonable for discussing actual data analysis.

These statistical errors already indicate that the damping time $\tau$ is not measured very precisely, because its error does not receive $Q^{-2}$ suppression. As Fig.~\ref{fig:qfactor} shows, $Q^2 >10$ for $\chi \gtrsim 0.686$, where $0.686$ results from mergers of equal-mass, nonspinning black holes \cite{2009PhRvD..79b4003S}. Thus, the frequency $f$ is determined precisely even for a single event with moderate $\rho$ thanks to $Q^{-2}$ suppression. Having said that, we recall that the statistical errors are suppressed further by taking catalog-level averages. This suppression enhances relative importance of the biases computed below.

Nondiagonal elements of $C^{ab}$ express the covariance and also play an important role. Again normalizing by $\rho^2$ before taking the average, we obtain
\begin{align}
 \frac{\ev{\rho^2 C^{A\phi}}}{A} & = - \frac{1}{4Q^3} - \frac{1}{16Q^5} , \\
 \frac{\ev{\rho^2 C^{Af}}}{Af} & = - \frac{1}{16Q^4} - \frac{1}{32Q^6} , \\
 \frac{\ev{\rho^2 C^{A\tau}}}{A\tau} & = -2 - \frac{3}{2Q^2} - \frac{5}{16Q^4} , \\
 \frac{\ev{\rho^2 C^{\phi f}}}{f} & = - \frac{1}{Q} - \frac{5}{4Q^3} - \frac{7}{32Q^5} , \\
 \frac{\ev{\rho^2 C^{\phi \tau}}}{\tau} & = \frac{7}{8Q^3} + \frac{3}{16Q^5} , \\
 \frac{\ev{\rho^2 C^{f\tau}}}{f\tau} & = - \frac{1}{4Q^4} .
\end{align}
As shown in App.~\ref{app:directave}, the $Q$-independent term of $C^{A\tau}$ is unchanged if we take the average before multiplying $\rho^2$. In fact, most of the $Q$-independent terms related to $A$ and $\tau$ are reproduced by the amplitude-only model with two parameters, $Ae^{-t/\tau}$ (the only exception is $\Delta \tau_\mathrm{ML}$ discussed in Sec.~\ref{sec:lg}, which is halved).

\section{Prior-gradient bias} \label{sec:pg}

In the Bayesian analysis commonly employed in gravitational-wave data analysis, the prior probability distribution $p (\theta^a)$ for waveform parameters needs to be specified. A nonflat prior tilts the posterior probability distribution and shifts, e.g., the Bayesian mean by (see, e.g., App.~A.~6 of Ref.~\cite{1994PhRvD..49.2658C})
\begin{equation}
 \delta \theta^a = C^{ab} \partial_b \ln p .
\end{equation}
Nonnegligible influence of the prior in black-hole spectroscopy is indeed mentioned in Ref.~\cite{2026arXiv260319021T}, and we scrutinize this shift in this section. Hereafter, we call this shift the prior-gradient bias. It should be cautioned that this naming relies on our presumably natural but specific parametrization of the waveform. If we adopt different parametrization, a part of this bias will be moved to the likelihood-geometry bias discussed in the next section.

Components of $\ev{\rho^2 C^{ab}}$ derived in the previous section imply that the damping time $\tau$ exhibits $Q$-independent prior-gradient biases when the prior on $A$ or $\tau$ is nonflat. Specifically, if their priors are given by a power-law form $p(A) \propto A^{-n_A}$ and $p (\tau) \propto \tau^{-n_\tau}$ in the relevant parameter range, the Bayesian mean of $\tau$ will be biased at $\order{\rho^{-2}}$ by
\begin{equation}
 \frac{\ev{\rho^2 \delta \tau}}{\tau} = 2n_A - 4n_\tau + \frac{3n_A - 3n_\tau}{2Q^2} + \frac{5n_A - 6n_\tau}{16Q^4} . \label{eq:pgtau}
\end{equation}
Intuitively, when the small amplitude is preferred by a large-$n_A$ prior, the signal needs to be compensated by the long damping time, and vice versa.

While there may be no specific reason to adopt a nonflat prior on $\tau$ when it is to be estimated, the prior on $A$ is frequently taken to be a decreasing function of $A$, i.e., $n_A>0$, producing a systematically positive deviation of $\tau$ from the Kerr hypothesis. In black-hole spectroscopy, a log-flat prior on $A$, which amounts to $n_A=1$, is a common choice \cite{2026arXiv260319021T}. A worse but still natural choice is $n_A=4$ for sources distributed uniformly in a Euclidean space. Then, this bias could accumulate in the catalog-level black-hole spectroscopy and eventually surpass statistical errors suppressed by $N^{-1/2}$ if the prior is not canceled. Because the relevant prior is not only for $\tau$ but for a kind of nuisance parameter $A$, naive combinations across the catalog may overlook this accumulation. The frequency $f$ also receives a relatively unsuppressed, $\order{Q^{-1}}$ bias via the prior on $\phi$ in principle, but this should be flat for most applications.

\begin{figure}[tbp]
 \includegraphics[width=.95\linewidth]{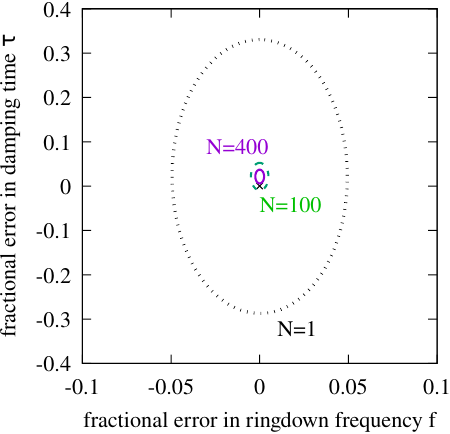}
 \caption{Contour enclosing $68.3\%$ probability of the fractional error from the true value marked by cross in the ringdown frequency $f$ and damping time $\tau$. The statistical error and the prior-gradient bias for $n_A=1$ is taken into account. The outer black-dotted ellipse denotes the error for a single event with $Q^2 = 10$ and $\rho = 10$. The middle green-dashed and inner purple-solid ellipses are results obtained by combining 100 and 400 events, respectively.} \label{fig:ellipse}
\end{figure}

Quantitatively, a few hundreds events can exhibit apparent violation of the Kerr hypothesis due to the prior-gradient bias. Figure \ref{fig:ellipse} shows the case of combining the posterior in a two-dimensional space spanned by $f$ and $\tau$, which may be directly relevant to the black-hole spectroscopy. Here, it is assumed that all the ringdown signals have $\rho=10$ and $Q^2=10$ in common and are analyzed with the log-flat-in-$A$ prior, $n_A=1$. The statistical error for a single event is (with a slight abuse of notation) $\ev{\sigma_\tau}/\tau \approx 0.2(\rho/10)^{-1}$ and is larger by an order of magnitude than the prior-gradient bias, $\ev{\delta \tau}/\tau \approx 0.02(\rho/10)^{-2}$. Because the former is suppressed approximately by $N^{-1/2}$, the Kerr hypothesis can be apparently violated for a catalog with $N \gtrsim 100(\rho/10)^2$.\footnote{If we consider uniformly distributed events in a Euclidean space with a detection threshold $\rho_*$, $\rho^2$ is replaced by $5\rho_*^2/3$. Considering that the gravitational-wave catalog has already reached significant cosmological redshifts, we avoid restricting our discussion to this simplistic distribution.} The actual number will be larger than this naive estimate because of the dimensionality, and the Kerr hypothesis is violated for $N \gtrsim 200$ in the case of Fig.~\ref{fig:ellipse}.

If we know the value of $n_A$, the $Q$-independent prior-gradient bias for $\tau$ will be canceled at the level of individual events by adopting $n_\tau = n_A/2$. For example, $n_A=1$ for the log-flat prior can be counteracted by $n_\tau=1/2$. Although this value has no astronomical motivation, this may be reasonably acceptable as a compromise between the flat, $n_\tau=0$ and log-flat, $n_\tau=1$ priors. If we do not have a direct control of $n_A$, detailed analysis including higher-dimensional Fisher analysis will be required to determine the appropriate value of $n_\tau$ or its equivalent to mitigate this bias.

\section{Likelihood-geometry bias} \label{sec:lg}

Even if the prior is flat, $\order{\rho^{-2}}$ biases are induced by the geometry of likelihood surface \cite{1994PhRvD..49.2658C,2008PhRvD..77d2001V}. Again, the damping time $\tau$ exhibits $Q$-independent positive biases under the perfect model and proper analysis. This type of biases, which we call the likelihood-geometry bias, can be suppressed by $N^{-1}$ in the catalog-level black-hole spectroscopy as long as the full individual likelihood for suitable parameters is appropriately combined in the Bayesian analysis (see Ref.~\cite{2017PhRvL.118p1101Y} for a related discussion). However, inappropriate joint inferences, e.g., combining only marginalized likelihood or point estimates, could accumulate this bias.

\subsection{Maximum-likelihood analysis} \label{sec:lg_ml}

First, we derive the likelihood-geometry bias of the maximum-likelihood estimator $\hat{\theta}_\mathrm{ML}^a$ \cite{1994PhRvD..49.2658C,2008PhRvD..77d2001V}. The $\order{\rho^{-1}}$ bias is linear in the noise realization $n(t)$, and thus its noise-ensemble average vanishes. The $\order{\rho^{-2}}$ bias is derived in Eq.~(57) of Ref.~\cite{2008PhRvD..77d2001V}, and its average over the noise ensemble is given by%\footnote{We believe that this is identical with Eq.~(59) of Ref.~\cite{2008PhRvD..77d2001V}.}
\begin{equation}
 \Delta \theta_\mathrm{ML}^a = - \frac{1}{2} C^{ad} C^{bc} (h_{bc}|h_d) ,
\end{equation}
where $h_{bc} \coloneqq \partial_b \partial_c h$. In this study, $\Delta \theta^a$ always represents noise-ensemble averaged biases. After scaling by $\rho^2$ and averaging over $\phi$, we obtain the explicit expressions
\begin{align}
 \frac{\ev{\rho^2 \Delta A_\mathrm{ML}}}{A} & = 1 + \frac{2}{Q^2} + \frac{27}{32Q^4} + \frac{3}{32Q^6} , \\
 \ev{\rho^2 \Delta \phi_\mathrm{ML}} & = \frac{1}{4Q^3} + \frac{1}{16Q^5} , \\
 \frac{\ev{\rho^2 \Delta f_\mathrm{ML}}}{f} & = - \frac{1}{8Q^4} - \frac{1}{16Q^6} , \\
 \frac{\ev{\rho^2 \Delta \tau_\mathrm{ML}}}{\tau} & = 4 + \frac{1}{2Q^2} .
\end{align}
We again recall that this averaging procedure is not unique. As shown in App.~\ref{app:directave}, the leading-in-$Q$ terms are unchanged except for the frequency, for which only the numerical factor is affected.

These results indicate that the bias of the damping time $\tau$ does not receive $Q^{-2}$ suppression as well as its statistical error. This is in stark contrast with the frequency $f$, which benefits from strong suppression by $Q^{-4}$. Quantitatively, for $\rho = 10$ and $Q^2 = 10$, $\ev{\Delta \tau_\mathrm{ML}} / \tau \approx 0.04 (\rho/10)^{-2}$ is larger by three orders of magnitude than $\ev{\Delta f_\mathrm{ML}} / f \approx -10^{-5} (\rho/10)^{-2}$. If this bias is accumulated due to an inappropriate combination across the catalog, $N \gtrsim \ev{\sigma_\tau^2}/\ev{\Delta \tau}^2 \approx 25(\rho/10)^2$ events could be sufficient for apparent violation of the Kerr hypothesis.

The likelihood-geometry bias of the damping time can be mitigated if we estimate the damping rate, $\gamma \coloneqq 1/\tau \, (= \omega_\mathrm{I})$, instead. Because this is a mere reparametrization, the statistical error is unchanged as $\ev{\sigma_\gamma^2}/\gamma^2 = \ev{\sigma_\tau}^2 / \tau^2$. However, the bias is modified by the square of the standard deviation. Specifically, the difference between the maximum-likelihood estimator and the true value before taking the ensemble average satisfies
\begin{align}
 \frac{\hat{\gamma}_\mathrm{ML} - \gamma}{\gamma} & = \pqty{1 + \frac{\hat{\tau}_\mathrm{ML} - \tau}{\tau}}^{-1} - 1 \\
 & \approx - \frac{\hat{\tau}_\mathrm{ML} - \tau}{\tau} + \pqty{\frac{\hat{\tau}_\mathrm{ML} - \tau}{\tau}}^2 ,
\end{align}
where we keep terms at the relevant order in the second line. This expression implies that
\begin{align}
 \frac{\ev{\rho^2 \Delta \gamma_\mathrm{ML}}}{\gamma} & = - \frac{\ev{\rho^2 \Delta \tau_\mathrm{ML}}}{\tau} + \frac{\ev{\rho^2 \sigma_\tau^2}}{\tau^2} \\
 & = \frac{1}{Q^2} + \frac{3}{8Q^4}
\end{align}
up to $\order{\rho^{-2}}$. This reparametrization postpones the apparent violation of the Kerr hypothesis until $N \approx \num{4e4}(\rho/10)^2$ events even for inappropriate joint inferences. Indeed, we may alternatively view that the $Q$-independent likelihood-geometry bias of $\tau$ is introduced by $\ev{\sigma_\gamma^2}/\gamma^2$. However, adopting $\gamma$ is not necessarily helpful if the Bayesian analysis is conducted.

\subsection{Bayesian analysis} \label{sec:lg_b}

Next, we derive the bias in the Bayesian posterior mean, $\hat{\theta}_\mathrm{B}^a$. This is given by Eq.~(73) of Ref.~\cite{2008PhRvD..77d2001V}, and its noise-ensemble average with the flat prior on all $\theta^a$ is given by
\begin{equation}
 \Delta \theta^a_\mathrm{B} = - C^{ab} C^{cd} [(h_b|h_{cd}) + (h_c|h_{bd})] .
\end{equation}
The explicit expressions for their $\phi$ averages are
\begin{align}
 \frac{\ev{\rho^2 \Delta A_\mathrm{B}}}{A} & = \frac{9}{2Q^2} + \frac{2}{Q^4} + \frac{7}{32Q^6} , \\
 \ev{\rho^2 \Delta \phi_\mathrm{B}} & = \frac{1}{2Q^3} + \frac{1}{16Q^5} , \\
 \frac{\ev{\rho^2 \Delta f_\mathrm{B}}}{f} & = - \frac{9}{16Q^4} - \frac{5}{32Q^6} , \\
 \frac{\ev{\rho^2 \Delta \tau_\mathrm{B}}}{\tau} & = 6 - \frac{3}{2Q^2} - \frac{5}{16Q^4} ,
\end{align}
where $\{A , \phi , f , \tau\}$ still denote the true value required for our predata Bayesian analysis. The Bayesian median is likely, although not necessarily, to be located between the Bayesian mean and the maximum-likelihood estimator.

Again, we find that the bias of the damping time does not receive $Q^{-2}$ suppression. Actually, the bias of the Bayesian mean is worse by a factor of $\approx 1.5$ than the maximum-likelihood bias. In Bayesian analysis, catalog events could be combined in multiple ways \cite{2021iSci...24j2577M,2024PhRvD.110d4053Z}, and appropriately joining the full likelihood with rescaled parameters will suppress the likelihood-geometry bias by $N^{-1}$. Still, naive joint inferences could lead to apparent violation of the Kerr hypothesis.

The prior-gradient bias discussed in Sec.~\ref{sec:pg} may be utilized to cancel the $Q$-independent likelihood-geometry bias of $\tau$ for individual events. While this can be controlled via the prior on the amplitude $A$ using the opposite approach from Sec.~\ref{sec:pg}, the prior on $\tau$ itself may be adjusted more flexibly. The power-law prior $p(\tau) \propto \tau^{-n_\tau}$ contributes to the bias by $\approx -4n_\tau$ [see Eq.~\eqref{eq:pgtau}], and thus the $Q$-independent likelihood-geometry bias is eliminated if we choose $n_\tau = 3/2$. Although this choice appears to have no astronomical motivation, this bias-canceling prior could be useful for some applications. If we would like to cancel both the prior-gradient and likelihood-geometry biases, we need to choose $n_\tau = (n_A+3)/2$. This cancellation, however, will be involved if $\tau$ is not directly chosen as a parameter.

Finally, we comment that the parametrization in terms of the damping rate is not advantageous in the Bayesian analysis, partly because the mean of $\gamma = 1/\tau$ is not the inverse of the mean of $\tau$, i.e., $\ev{\gamma} = \ev{1/\tau} \neq 1/\ev{\tau}$. Quantitatively, the flat prior on $\gamma$ derives
\begin{equation}
 \frac{\ev{\rho^2 \Delta \gamma_\mathrm{B}}}{\gamma} = 10 + \frac{15}{2Q^2} + \frac{29}{16Q^4} ,
\end{equation}
leading to a stronger bias than that for $\tau$. Still, we may be able to find better parametrization by, e.g., geometric considerations \cite{2003PhRvD..68j2003N}. We leave this investigation as a future task.

\section{Summary and discussion} \label{sec:summary}

In this study, we investigate biases caused by the finite signal-to-noise ratio in parameter estimation for damped oscillations as the simplest model of the black-hole spectroscopy. We find that the damping time $\tau$ suffers from two types of plausibly positive biases that are not suppressed by the quality factor $Q = \pi f \tau$ even if the ringdown model is perfect and the analysis is properly conducted. One is the prior-gradient bias, which is likely positive because of the typically decreasing prior on the amplitude. The other is the likelihood-geometry bias, which is also positive. Although both can be suppressed in the catalog-level black-hole spectroscopy if the likelihood is carefully combined, it may not be straightforward in actual implementations. Therefore, combining $\gtrsim 100$ events with $\rho \approx 10$ without due care could lead to apparent violation of the Kerr hypothesis. If the catalog-level black-hole spectroscopy would find that the damping time is significantly longer than the theoretical expectation, its interpretation requires particular care.

We propose simple strategies for mitigating these biases at the level of individual events. If we know the prior on the amplitude, the $Q$-independent prior-gradient bias of $\tau$ can be counteracted by adjusting its prior. The likelihood-geometry bias can also be mitigated. If the maximum-likelihood estimator is considered in frequentist analysis, the damping rate $\gamma = 1/\tau$ is a preferable parametrization to $\tau$, because its bias receives $Q^{-2}$ suppression. However, this does not mitigate the bias in the Bayesian analysis. Instead, again adjusting the prior counteracts the $Q$-independent bias.

Obviously, the finite signal-to-noise ratio bias is merely one of many possible biases. Given the strong belief in the Kerr black hole as the merger remnant, comprehensive investigations are necessary before drawing a strong conclusion from the black-hole spectroscopy if any violation would seem to be found. In particular, these investigations should adopt more realistic waveform models and analysis pipelines. Although all the biases studied here are trivially circumvented by focusing only on high-$\rho$ ringdown events, as partly consistent with the result reported in Ref.~\cite{2026arXiv260319021T}, various biases caused by mismodeling will generally become prominent for these high-$\rho$ events. Thus, there will always be a compromise. We believe that the finite signal-to-noise ratio biases derived in this study are so robust that they need to be taken into account in a wide class of black-hole spectroscopy.

\begin{acknowledgements}
 We thank Hiroyuki Nakano, Naoki Seto, and Nami Uchikata for helpful comments. This work was supported by JSPS KAKENHI Grant-in-Aid for Scientific Research (No.~JP26K07062).
\end{acknowledgements}

\appendix

\section{Average without normalizing by the signal-to-noise ratio} \label{app:directave}

We present the average over $\phi$ of various quantities without normalizing by $\rho$. We instead multiply $\ev{\rho^2}$ after averaging to keep the appearance similar to the counterparts presented in the main text. The statistical errors are
\begin{align}
 \ev{\rho^2} \frac{\ev{\sigma_A^2}}{A^2} & = 2 + \frac{2}{Q^2} + \frac{5}{8Q^4} + \frac{1}{16Q^6} , \\
 \ev{\rho^2} \ev{\sigma_\phi^2} & = 2 + \frac{2}{Q^2} + \frac{5}{8Q^4} + \frac{1}{16Q^6} , \\
 \ev{\rho^2} \frac{\ev{\sigma_f^2}}{f^2} & = \frac{1}{Q^2} + \frac{1}{2Q^4} + \frac{1}{16Q^6} , \\
 \ev{\rho^2} \frac{\ev{\sigma_\tau^2}}{\tau^2} & = 4 + \frac{2}{Q^2} + \frac{1}{4Q^4} ,
\end{align}
and nondiagonal elements of the covariance matrix are
\begin{align}
 \ev{\rho^2} \frac{\ev{C^{A\phi}}}{A} & = 0 , \\
 \ev{\rho^2} \frac{\ev{C^{Af}}}{Af} & = - \frac{1}{4Q^4} - \frac{1}{16Q^6} , \\
 \ev{\rho^2} \frac{\ev{C^{A\tau}}}{A\tau} & = -2 - \frac{2}{Q^2} - \frac{3}{8Q^4} , \\
 \ev{\rho^2} \frac{\ev{C^{\phi f}}}{f} & = - \frac{1}{Q} - \frac{1}{Q^3} - \frac{3}{16Q^5} , \\
 \ev{\rho^2} \frac{\ev{C^{\phi \tau}}}{\tau} & = \frac{1}{2Q^3} + \frac{1}{8Q^5} , \\
 \ev{\rho^2} \frac{\ev{C^{f\tau}}}{f\tau} & = 0 .
\end{align}
The biases of the maximum-likelihood estimators are
\begin{align}
 \ev{\rho^2} \frac{\ev{\Delta A_\mathrm{ML}}}{A} & = 1 + \frac{2}{Q^2} + \frac{15}{16Q^4} + \frac{1}{8Q^6} , \\
 \ev{\rho^2} \ev{\Delta \phi_\mathrm{ML}} & = \frac{1}{4Q^3} + \frac{1}{16Q^5} , \\
 \ev{\rho^2} \frac{\ev{\Delta f_\mathrm{ML}}}{f} & = - \frac{1}{4Q^4} - \frac{1}{16Q^6} , \\
 \ev{\rho^2} \frac{\ev{\Delta \tau_\mathrm{ML}}}{\tau} & = 4 + \frac{1}{Q^2} ,
\end{align}
and those of the Bayesian posterior mean are
\begin{align}
 \ev{\rho^2} \frac{\ev{\Delta A_\mathrm{B}}}{A} & = \frac{4}{Q^2} + \frac{9}{4Q^4} + \frac{5}{16Q^6} , \\
 \ev{\rho^2} \ev{\Delta \phi_\mathrm{B}} & = \frac{1}{2Q^3} + \frac{1}{8Q^5} , \\
 \ev{\rho^2} \frac{\ev{\Delta f_\mathrm{B}}}{f} & = - \frac{3}{4Q^4} - \frac{3}{16Q^6} , \\
 \ev{\rho^2} \frac{\ev{\Delta \tau_\mathrm{B}}}{\tau} & = 6 - \frac{3}{8Q^4} ,
\end{align}
Detailed computations required to derive these results are presented in an accompanying Mathematica file, where essentially the same Fisher-matrix components have been presented in Ref.~\cite{2006PhRvD..73f4030B}.

%\bibliography{paper}
%
\end{document}